\documentclass[a4paper]{report}
\usepackage[utf8]{inputenc}
\usepackage[T1]{fontenc}
\usepackage{RJournal}
\usepackage{booktabs}
\usepackage{array, amsthm, amsmath, amssymb, mathrsfs, multirow, url, subfigure}
\usepackage{graphicx} 
\usepackage{amsfonts}
\usepackage{algorithm, algorithmicx, algpseudocode, gensymb}

\graphicspath{{./figures/}}

\newcolumntype{L}{>$c<$}
\newcolumntype{R}{>$c<$}
\theoremstyle{plain}

\theoremstyle{definition}

\theoremstyle{remark}

\newcommand{\unif}{{\sf Unif}}
\newcommand{\nm}{{\sf N}}

\newcommand{\gam}{{\sf Gamma}}

\newcommand{\veta}{\boldsymbol{\veta}}

\renewcommand{\phi}{\varphi}

\begin{document}

%% do not edit, for illustration only
\sectionhead{Contributed research article}
\volume{XX}
\volnumber{YY}
\year{20ZZ}
\month{AAAA}

%% replace RJtemplate with your article
\begin{article}
% !TeX root = RJwrapper.tex
\title{BayesBD: An R Package for Bayesian Inference on Image Boundaries}
\author{by Nicholas Syring and Meng Li}

\maketitle

\abstract{
	
	We present the \pkg{BayesBD} package providing Bayesian inference for boundaries of noisy images.  The \pkg{BayesBD} package implements flexible Gaussian process priors indexed by the circle to recover the boundary in a binary or Gaussian noised image, with the benefits of guaranteed geometric restrictions on the estimated boundary, (nearly) minimax optimal and smoothness adaptive convergence rates, and convenient joint inferences under certain assumptions. The core sampling tasks for our model have linear complexity, and our implementation in \code{c++} using packages \pkg{Rcpp} and \pkg{RcppArmadillo} is computationally efficient. Users can access the full functionality of the package in both \code{Rgui} and the corresponding \pkg{shiny} application. Additionally, the package includes numerous utility functions to aid users in data preparation and analysis of results. We compare \pkg{BayesBD} with selected existing packages using both simulations and real data applications, and demonstrate the excellent performance and flexibility of \pkg{BayesBD} even when the observation contains complicated structural information that may violate its assumptions. }

\section{Introduction} 
Boundary estimation is an important problem in image analysis with wide-ranging applications from identifying tumors in medical images~\citep{Li.2010}, classifying the process of machine wear by analyzing the boundary between normal and worn materials~\citep{yuan.2016} to identifying regions of interest in satellite images such as to detect the boundary of Lake Menteith in Scotland from a satellite image~\citep{cucala, marin_book}. Furthermore, boundaries present in epidemiological or ecological data may reflect the progression of a disease or an invasive species; see~\citet{Waller+Gotway:04}, \citet{Lu+Car:05}, and \citet{Fit+:10}. 

% Dense spatial data may also be analyzed using pixel-level methods. 
There is a rich literature about image segmentation for both noise-free and noisy observations; see the surveys in \citet{Ziou.Tabbone.1998, Basu:02, Maini.Aggarwal.2009, Bha+Mit:12}. For Bayesian approaches to image segmentation, see \citet{bayes_image_book} and \citet{gren_book}. 
Recently, \citet{Li.Ghosal.2015} developed a flexible nonparametric Bayesian model to detect image boundaries, which achieved four aims of guaranteed geometric restriction, (nearly) minimax optimal rate adaptive to the smoothness level, convenience for joint inference, and computational efficiency. However, despite the theoretical soundness, the practical implementation of Li and Ghosal’s method is far from trivial, mostly in the approachability of the proposed nonparametric Bayesian framework and further improvement in the speed of posterior sampling algorithms, which becomes critical in attempts to popularize this approach in statistics and the broader scientific community. In this paper, we present the \code{R} package \CRANpkg{BayesBD} \citep{Syring.Li.BD.2016} which aims to fill this gap. The developed \pkg{BayesBD} package provides support for analyzing binary images and Gaussian-noised images, which commonly arise in many applications. We implement various options for posterior calculation including the Metropolis-Hastings sampler~\citep{MH} and slice sampler~\citep{neal.2003}. To further speed up the Markov Chain Monte Carlo (MCMC), we take advantage of the integration via \CRANpkg{RcppArmadillo}~\citep{Rcpp, RcppArmadillo} of \code{R} and the compiled \code{c++} language. We further integrate the \pkg{BayesBD} package with \CRANpkg{shiny}~\citep{shiny} to facilitate the usage of implemented boundary detection methods in real applications. 

A far as we know, there are no other \code{R} packages for image boundary detection problems achieving the four goals mentioned above. An earlier version of the \pkg{BayesBD} package~\citep{Li.BD.2015}
provided first-of-its-kind tools for analyzing images, but support for Gaussian-noised images, \code{c++} implementations, more choices of posterior samplers, and \pkg{shiny} integrations were not available until the current version. For example, the nested loops required for MCMC sampling were inefficient in \code{R} programming. The combination of new programming and faster sampling algorithms means that a typical simulation example consisting of $5000$ posterior samples from $10000$ data points can now be completed in about one minute.  

%was released through CRAN, \citet{R} 

The rest of the paper is organized as follows. We first introduce the problem of statistical inference on boundaries of noisy images, the nonparametric Bayesian models in use, and posterior sampling algorithms. We then demonstrate how to use the main functions of the package for data analysis working with both \code{Rgui} and \pkg{shiny}.  We next conduct a comprehensive experiment on the comparison of sampling methods and coding platforms, scalability of \pkg{BayesBD}, and comparisons with existing packages including \pkg{mritc}, \pkg{bayesImageS}, and \pkg{bayess}.  We illustrate a pair of real data analyses of medical and satellite images. The paper is concluded by a Summary section. 

\section{Statistical analysis of image boundaries}
\label{S:2}
\subsection{Image data}
\begin{figure}[!ht]
	\centering
	\includegraphics[width=0.4\textwidth, trim={2cm 2cm 2cm 2cm},clip]{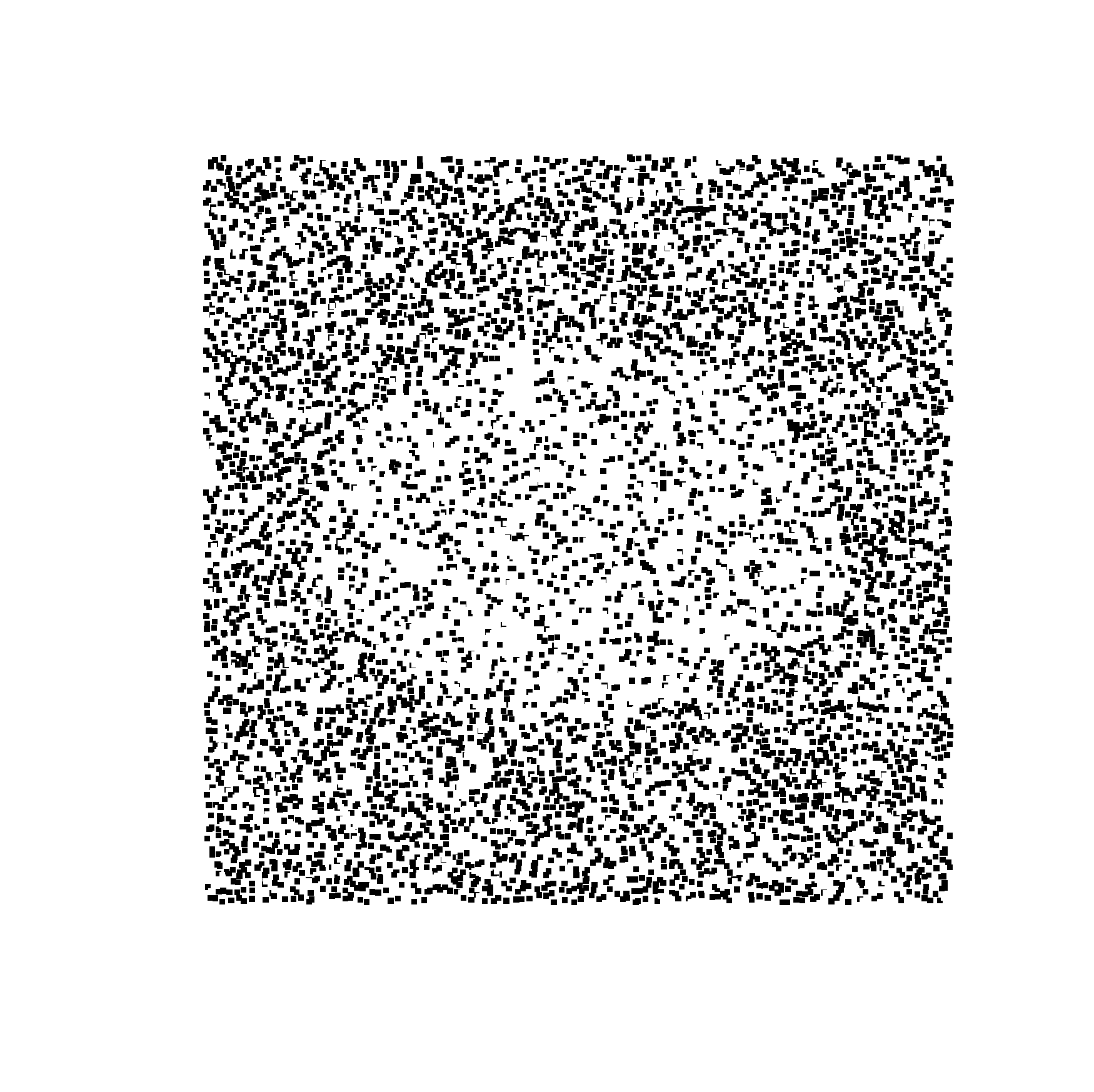}
	\includegraphics[width=0.4\textwidth, trim={2cm 2cm 2cm 2cm},clip]{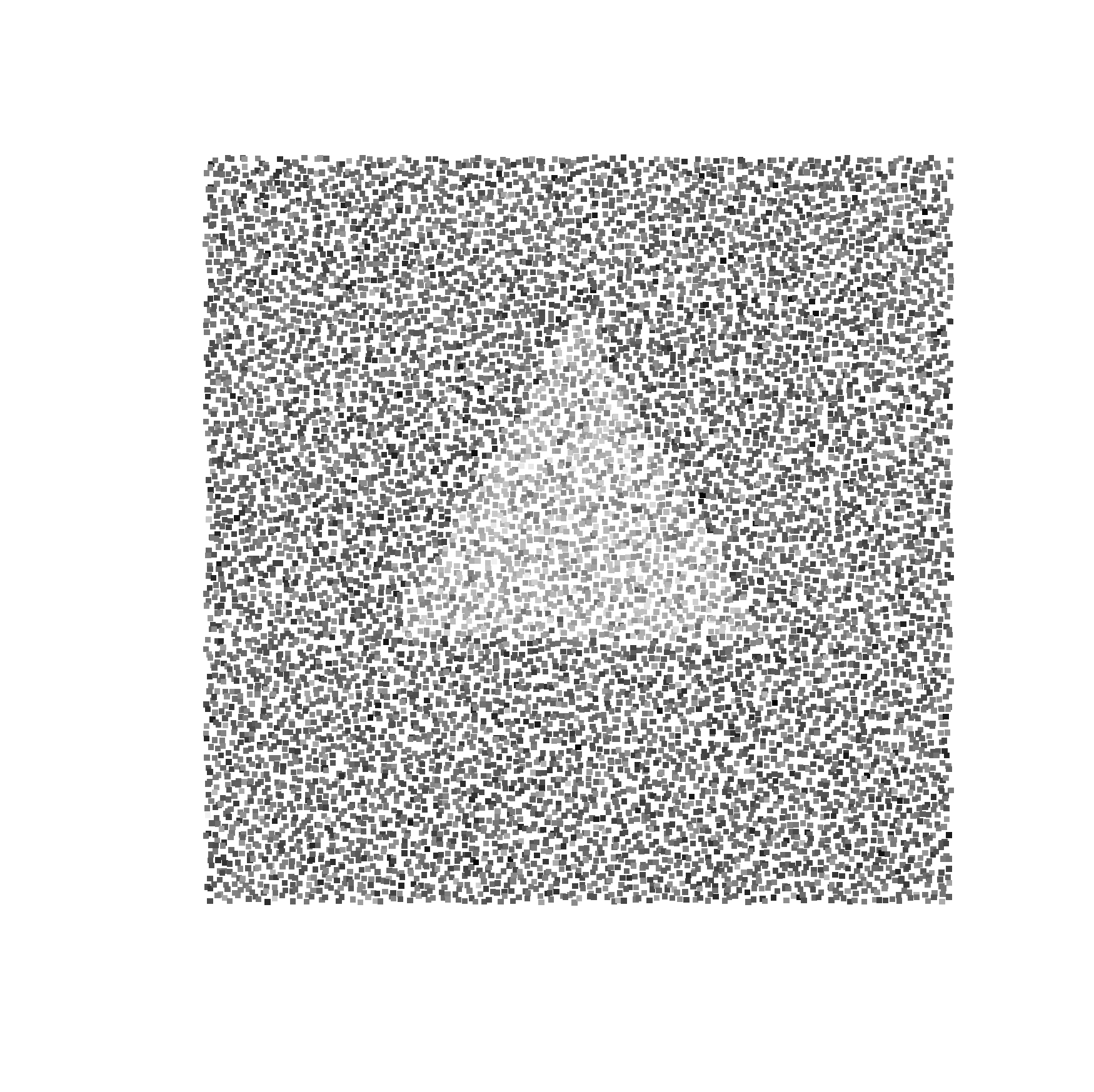}
	\caption{Left: a binary image generated using an elliptical boundary and parameters $\pi_1 = 0.65$ and $\pi_2 = 0.35$.  Right: a Gaussian-noised image generated using a triangular boundary and parameters $\mu_1 = 1$, $\mu_2 = -1$, and $\sigma_1 = \sigma_2 = 1$. Both images have the size $100 \times 100$. }
	\label{fig:example_data}
\end{figure}

An image observed with noise may be represented by a set of data points or pixels $(Y_i, X_i)_{i=1}^n$, where the intensities $Y_i$ are observed at locations $X_i \in \mathcal{X} = [0,1]^2$. Following~\cite{Li.Ghosal.2015}, we assume that there is a closed region $\Gamma \in \mathcal{X}$ such that the intensities $Y_i$ are distributed as follows conditionally on whether its location is inside $\Gamma$ or in the background $\Gamma^c$:
\begin{equation}
Y_i \sim \left \{
\begin{array}{cc}
f(\cdot; \xi) & X_i \in \Gamma;\\
f(\cdot; \rho) & X_i \in \Gamma^c, \label{eq:setup}
\end{array} \right. 
\end{equation}    
where $f$ is a given probability mass function or probability density function of a parametric family up to unknown parameters $(\xi, \rho)$. For example, Figure~\ref{fig:example_data} shows two simulated images where the parametric family is Bernoulli and Gaussian, respectively. These images can be reproduced using the functions \code{par2obs}, \code{parnormobs}, and \code{plotBD} which will be  demonstrated in detail later on. 

The parameter of interest is the boundary of the closed region $\gamma := \partial \Gamma$, which is assumed to be closed and smooth, while $(\xi, \rho)$ are nuisance parameters. We make the following assumptions about the noisy image:
\begin{enumerate}
	\item The pixel locations $X_i$ are sampled either completely randomly, i.e., $X_i \stackrel{i.i.d.}{\sim}\unif(\mathcal{X} )$ or jitteredly randomly, i.e., $\mathcal{X}$ is first partitioned into blocks $\mathcal{X}_i$ using an equally-spaced grid and then locations are sampled $X_i\sim \unif(\mathcal{X}_i)$.
	\item The closed region $\Gamma$ is star-shaped with a known reference point $O \in \Gamma$, i.e., the line segment joining $O$ to any point in $\Gamma$ is also in $\Gamma$.  
	\item The boundary $\gamma$ is an $\alpha-$H\"older smooth function, i.e., $\gamma \in \mathbb{C}^\alpha(\mathbb{S})$ where
	\[\mathbb{C}^\alpha(\mathbb{S}) := \{f: \mathbb{S}\rightarrow \mathbb{R}^+, |f^{(\alpha_0)}(x) - f^{(\alpha_0)}(y)|\leq L_f\|x-y\|^{\alpha - \alpha_0},\, \forall x,y \in \mathbb{S}\}\]
	where $\mathbb{S}$ is the unit circle, $\alpha_0$ is the largest integer strictly smaller than $\alpha$, $L_f$ is some positive constant and $\|\cdot\|$ is the Euclidean distance.
\end{enumerate}

Assumptions 2 and 3 imply that the region of interest is star-shaped with a smooth boundary. While these assumptions are crucial to guarantee desirable asymptotic properties of the estimator implemented in \pkg{BayesBD} and are reasonable in many applications, it is certainly of great interest to investigate the performance of \pkg{BayesBD} when these assumptions are violated. In what follows, we study numerous examples that are not uncommon in practice but violate these assumptions to some extent, to demonstrate the flexibility of \pkg{BayesBD} and its capacity to handle practical images that may be much more complicated than the two-region setting with assumptions above. These examples include the triangular boundary in Figure~\ref{fig:example_data} which has a piecewise smooth boundary. and thus violates Assumption 3, and three real data examples including the image of Lake Menteith~\ref{fig:compare_lake} and two neuroimaging examples~\ref{fig:pet} where multiple regions are present and the region of interest has non-smooth or even discontinuous boundary.

Letting $\Theta$ be the parameter space of the parametric family $f$, conditions to separate the inside and outside parameters are needed. Examples of the parameter space $\Theta^*$ for $(\xi, \rho)$ include but are not limited to:
\begin{itemize}
	\item[4A.] One-parameter family such as Bernoulli, Poisson, exponential distributions, and  $\Theta^* = \Theta^2 \cap \{(\xi, \rho): \rho < \xi \}$, or $\Theta^* = \Theta^2 \cap \{(\xi, \rho): \rho > \xi \}$.
	\item[4B.] Two-parameter family such as Gaussian distributions, and $\Theta^* = \Theta^2 \cap \{((\mu_1, \sigma_1), (\mu_2, \sigma_2)): \mu_1 > \mu_2, \sigma_1 = \sigma_2 \}$, or $\Theta^* = \Theta^2 \cap \{((\mu_1, \sigma_1), (\mu_2, \sigma_2)): \mu_1 > \mu_2, \sigma_1 > \sigma_2 \}$, or $\Theta^* = \Theta^2 \cap \{((\mu_1, \sigma_1), (\mu_2, \sigma_2)): \mu_1 = \mu_2, \sigma_1 > \sigma_2 \}$.
\end{itemize}
In practice, the order restriction in 4A or 4B is often naturally obtained depending on the concrete problem. For instance, in brain oncology, a tumor often has higher intensity values than its surroundings in a positron emission tomography scan, while for astronomical applications objects of interest emit light and will be brighter. A more general condition for any parametric family can be referred to Condition (C) in~\cite{Li.Ghosal.2015}.  

It is worth noticing that although model~\ref{eq:setup} follows a two-region framework, the method in~\cite{Li.Ghosal.2015} and our developed \pkg{BayesBD} have the flexibility to handle data with multiple regions by running the two-region method iteratively, which is demonstrated in the neuroimaging application below. 
\subsection{A nonparametric Bayesian model for image boundaries}
Let $Y = \{Y_i\}_{i = 1}^n$ and $X = \{X_i\}_{i = 1}^n$, then the likelihood of the image data described in~\eqref{eq:setup} is 
\[L(Y|X,\theta) = \prod_{i\in I_1}f(Y_i;\xi)\prod_{i \in I_2}f(Y_i;\rho),\]
where $I_1 = \{i:X_i \in \Gamma\}$, $I_2 = \{i:X_i \in \Gamma^c\}$, and $\theta$ denotes the full parameter $(\xi, \rho, \gamma)$.  

We view $\gamma$ as a curve mapping $[0,2\pi] \rightarrow \mathbb{R}^{+}$ and model it using a randomly rescaled Gaussian process prior on the circle $\mathbb{S}$: $\gamma(\omega) \sim \mathsf{GP}(\mu(\omega), G_a(\cdot, \cdot)/\tau)$ where the covariance kernel
\begin{eqnarray*}
	G_a(t_1, t_2) &=&  
	\exp({-a^2 \{(\cos 2\pi t_1 - \cos 2 \pi t_2)^2 + (\sin 2\pi t_1 - \sin 2\pi t_2)^2 \} }) \\
	&=& \exp\{- 4a^2 \sin^2(\pi t_1 - \pi t_2) \}
\end{eqnarray*}
is the so-called {\em squared exponential periodic kernel} obtained by mapping the squared exponential kernel on unit interval $[0, 1]$ to the circle through $Q: [0,1] \rightarrow \mathbb{S}, \;\omega \rightarrow (\cos 2 \pi \omega, \sin 2 \pi \omega)$ as in~\citet{MacKay:98}. The parameters $a$ and $\tau$ control the smoothness and scale of the kernel, respectively.  As shown in~\citet{Li.Ghosal.2015}, the covariance kernel has the following closed form decomposition: $G_a(t,t') = \sum_{k=1}^\infty v_k(a)\psi_k(t)\psi_k(t')$ where 
\[v_1(a) = e^{-2a^2}I_0(2a^2), \quad v_{2j}(a) = v_{2j+1}(a) = e^{-2a^2}I_j(2a^2), \; j\geq 1,\] and $I_n(x)$ denotes the modified Bessel function of the first kind of order $n$ and $\psi_j(t)$ is the $j$th Fourier basis function in $\{1, \cos 2\pi t, \sin 2\pi t,...\}$.  The above expansion allows us to write the boundary as a sum of basis functions: 
\begin{equation}
\label{eq:decomposition} 
\gamma(\omega) = \mu(\omega) + \sum_{k=1}^\infty z_k\psi_k(\omega), 
\end{equation} where $z_k \sim \nm(0, v_k(a)/\tau)$.  In practice, we truncate this basis function expansion using the first $L$ functions, i.e., $\gamma(\omega)  = \mu(\omega)+\sum_{k=1}^L z_k\psi_k(\omega)$.  In the \pkg{BayesBD} package, we use $L = 2J + 1$ with the default $J = 10$, which seems adequate for accurate approximation of $\gamma(w)$ as shown in \citet{Li.Ghosal.2015}, but users may specify a different value depending on the application.  

We use a Gamma prior distribution Gamma($\alpha_a, \beta_a$) for the rescaling factor $a$. This random rescaling scheme is critical to obtain rate adaptive estimates without assuming the smoothness level $\alpha$ in Assumption 3 is known; see, for example, ~\citet{van+van:09} and~\citet{Li.Ghosal.2015}. The default values of hyperparameters are $\alpha_a = 2$ and $ \beta_a = 1$. 

We use a constant function as the prior mean function $\mu(\cdot)$, with value determined by user input or by an initial maximum likelihood estimation.  The other hyperparameter and parameters follow standard conjugate priors. Specifically, we use a Gamma distribution Gamma($\alpha_{\tau}, \beta_{\tau}$) prior for $\tau$ with default values $\alpha_{\tau} = 500$ and $\beta_{\tau} = 1$. Priors for the nuisance parameters $\xi$ and $\rho$ depend on the parametric family $f$, which are 
\begin{itemize}
	\item Binary images: the parameters are the probabilities $(\pi_1, \pi_2) \sim \mathrm{OIB}(\alpha_1, \beta_1, \alpha_1, \beta_1)$, where OIB stands for ordered independent Beta distributions. 
	\item Gaussian noise: the parameters are the mean and standard deviation $(\mu_1, \sigma_1, \mu_2, \sigma_2)$ with prior distributions $ (\mu_1, \mu_2) \sim \mathrm{OIN}(\mu_0, \sigma_0^2, \mu_0, \sigma_0^2)$ and $(\sigma_1^{-2}, \sigma_2^{-2}) \sim \mathrm{OIG}(\alpha_2, \beta_2,  \alpha_2, \beta_2)$, where OIN and OIG are ordered independent normal and Gamma distributions, respectively. 
\end{itemize}
The orders in OIB, OIN and OIG are provided by users if such information is available; otherwise, the ordered independent distributions revert to independent distributions. Our default specifications are chosen to make the corresponding prior distributions spread out. For example, in the \pkg{BayesBD} package, the default values are $\alpha_1 = \beta_1 = 0$ for binary images, and $\mu_0 = \bar{Y}, \sigma_0 = 10^3$ and $\alpha_2 = \beta_2 = 10^{-2}$ for Gaussian noise, where $\bar{Y}$ is the same mean of all intensities. Under Assumptions 1--4, ~\citet{Li.Ghosal.2015} proved that the nonparametric Bayes approach is (nearly) rate-optimal in the minimax sense, adaptive to unknown smoothness level $\alpha$.

\subsection{Posterior sampling and estimation of the boundary}
Let $z = \{z_i\}_{i = 1}^L$ and $\Sigma_a = \text{diag}(v_1(a), \ldots, v_L(a))$. We use Metropolis-Hastings (MH) with the Gibbs sampler~\citep{geman.1984} to sample the joint distribution of the parameters $(z, \xi, \rho, \tau, a)$, where the MH step is for the vector parameter $z$. We also allow a slice sampling with the Gibbs sampler where slice sampling is used for $z$ as in~\cite{Li.Ghosal.2015}. We give the detailed sampling algorithms for binary image in Algorithm~\ref{algo:sample} and Gaussian-noised images in Algorithm~\ref{algo:sample2}. Comparisons between MH and slice sampling, along with other numerical performances are referred to the below section on \strong{Performance tests}. 
\begin{algorithm*}[!ht]
	\smallskip
	Initialize the parameters: $z = 0$, $\tau = 500$, $a = 1$, and $\mu(\cdot)$ is taken to be constant, i.e. a circle.  Then, initialize $(\xi, \rho) = (\pi_1, \pi_2)$ by the maximum likelihood estimates given $\mu(\cdot)$.    
	\begin{enumerate}
		\item At iteration $t+1$, sample $z^{(t+1)}|(\pi_1^{(t)}, \pi_2^{(t)},\tau^{(t)},a^{(t)},Y,X)$ one entry at a time, using either MH sampling and slice sampling, using the following logarithm of the conditional posterior density
		\[N_1 \log\frac{\pi_1^{(t)}(1-\pi_2^{(t)})}{\pi_2^{(t)}(1-\pi_1^{(t)})} + n_1\log\frac{1-\pi_1^{(t)}}{1-\pi_2^{(t)}}-\frac{\tau}{2} (z^{(t)})^\top \Sigma_{a^{(t)}}^{-1}z^{(t)},\]
		where $n_1 = \sum_{i = 1}^n 1(r_i < \gamma_i^{(t)})$ and $N_1 = \sum_{i = 1}^n 1(r_i < \gamma_i^{(t)})Y_i$; here 
		$(r_i,\omega_i)$ are the polar coordinates of pixel location $X_i$ and $\gamma_i^{(t)} \:= {\gamma}^{(t)}(\omega_i)$ is the radius of the image boundary at iteration $t$ and the $i$th pixel.
		\item Sample $\tau^{(t+1)}|z^{(t+1)},a^{(t)}\sim \gam(\alpha^\star,\beta^\star)$ where $\alpha^\star = \alpha_{\tau} + L/2$ and $\beta^\star = \beta_{\tau} + (z^{(t+1)})^\top\Sigma^{-1}_{a^{(t)}}z^{(t+1)}/2$.
		\item Sample $(\pi_1, \pi_2)|(z,Y)\sim \sf{}{OIB}$$( \alpha_1+N_1,\beta_1+n_1-N_1,\alpha_1+N_2,\beta_1+n_2-N_2)$, where $n_2 = \sum_{i = 1}^n 1(r_i \geq \gamma_i^{(t)})$ and $N_2 = \sum_{i = 1}^n 1(r_i \geq \gamma_i^{(t)})Y_i$. 
		\item Sample $a^{(t+1)}|(z^{(t+1)}, \tau^{(t+1)})$ by slice sampling using the logarithm of the conditional posterior density
		\[-\sum_{k=1}^L \frac{\log v_k(a^{(t)})}{2}-\sum_{k=1}^L\frac{\tau^{(t+1)}z_k^2}{2v_k(a^{(t)})}+(\alpha_a - 1)\log a^{(t)} - \beta_{a}.\]
	\end{enumerate}
	\caption{\bf-- Binary images.} 
	\label{algo:sample}
\end{algorithm*}

\begin{algorithm*}[!ht] 
	\smallskip
	Initialize the parameters: $z = 0$, $\tau = 500$, $a = 1$, and $\mu(\cdot)$ is taken to be constant, i.e. a circle.  Then, initialize $(\xi, \rho) = (\pi_1, \pi_2)$ by the maximum likelihood estimates given $\mu(\cdot)$.    
	\begin{enumerate}
		\item At iteration $t+1$, sample $z^{(t+1)}|(\mu_1^{(t)},\sigma_1^{(t)}, \mu_2^{(t)}, \sigma_2^{(t)},\tau^{(t)},a^{(t)},Y,X)$ one entry at a time, using either slice sampling or MH sampling, using the following logarithm of the conditional posterior density
		\[-n_1(\log \sigma_1^{(t)} - \log \sigma_2^{(t)})-\sum_{i \in I_1}\frac{(Y_i - \mu_1^{(t)})^2}{2(\sigma_1^2)^{(t)}}-\sum_{i \in I_2}\frac{(Y_i - \mu_2^{(t)})^2}{2(\sigma_2^2)^{(t)}}-\frac{\tau (z^{(t)})^\top \Sigma_{a^{(t)}}^{-1}z^{(t)}}{2}.\]
		\item Sample $\tau^{(t+1)}|z^{(t+1)},a^{(t)}$ as in Algorithm~1.
		\item Sample $(\mu_1, \sigma_1, \mu_2, \sigma_2) | (z,Y)$ conjugately.  
		\begin{itemize}
			\item Sample $(\sigma_1^{-2})^{(t+1)}$ from a Gamma distribution with parameters
			\[\alpha = \alpha_2+\frac{n_1}{2}, \hspace{1cm} \beta = \beta_2+\sum_{i \in I_1}\frac{(Y_i - \bar{Y}^{(1)})^2}{2}+\frac{\sigma_0^{-2}n_1}{n_1+\sigma_0^{-2}}\frac{(\bar{Y}^{(1)} - \mu_0)^2}{2},\]
			where $\bar{Y}^{(1)}$ is the sample mean of intensities in $I_1$. 
			\item sample $\mu_1^{(t+1)}$ from a normal distribution with mean and standard deviation
			\[\frac{\sigma_0^{-2}\mu_0}{n_1+\sigma_0^{-2}} + \frac{n_1\bar{y}_1}{n_1+\sigma_0^{-2}}, \hspace{1cm} (n_1+\sigma_0^{-2})^{-1/2}.\]
			\item Sample $(\sigma_2^{-2})^{(t+1)}$ and $\mu_2^{(t+1)}$ analogously.
			\item If \texttt{ordering} information is available, sort $(\mu_1^{(t + 1)}, \mu_2^{(t+1)})$ and 
			$(\sigma_1^{(t + 1)}, \sigma_2^{(t+1)})$ accordingly.
		\end{itemize}
		\item Sample $a^{(t+1)}|(z^{(t+1)}, \tau^{(t+1)})$ as in Algorithm~1.
	\end{enumerate}
	\caption{\bf-- Gaussian-noised images.}
	\label{algo:sample2}
\end{algorithm*}

Let $\{{\gamma}_t(\omega)\}_{t = 1}^T$ be the posterior samples after burn-in where $T$ is the number of posterior samples. We use the posterior mean as the estimate and construct a variable-width uniform credible band. Specifically,  let $(\widehat{\gamma}(\omega), \widehat{s}(\omega))$ 
be the posterior mean and standard deviation functions derived from $\{{\gamma}_t(\omega)\}$. For the $t$th MCMC run, we calculate the distance $u_t = \|(\gamma_t -\widehat{\gamma})/s\|_\infty=\sup_\omega\{ |\gamma_t(\omega) - \widehat{\gamma}(\omega)| /\widehat{s}(\omega) \}$ and obtain the 95th percentile of all the $u_t$'s, denoted as $L_0$. Then a 95\% uniform credible band is given by $[\widehat{\gamma}(\omega) - L_0 \widehat{s}(\omega), \widehat{\gamma}(\omega) + L_0 \widehat{s}(\omega)]$.

\section{Analysis of image boundaries using BayesBD from Rgui}
\label{S:3}
There are three steps to our Bayesian image boundary analysis: load the image data into \code{R} in the appropriate format, use the functions provided to sample from the joint posterior of the full parameter $\theta = (\xi, \rho, \gamma)$, and summarize the posterior samples both numerically and graphically.

\subsection{Generating image data} 
\label{sec:simulate.data} 
Two functions are included in \pkg{BayesBD} to facilitate data simulation for numerical experiments: \code{par2obs} for binary images and \code{parnormobs} for Gaussian-noised images.  Table~\ref{tbl:par2obs} describes the function arguments to \code{par2obs}, which returns sampled intensities and pixel locations in both Euclidean and polar coordinates. The function \code{parnormobs} is similar, with the replacement of arguments \code{pi.in} and \code{pi.out} by \code{mu.in}, \code{mu.out}, \code{sd.in}, and \code{sd.out} corresponding to parameters $\mu_1$, $\mu_2$, $\sigma_1$, and $\sigma_2$, respectively.

As a demonstration, the following code generates a $100 \times 100$ binary image of an ellipse using a jitteredly-random design with a reference point of (0.5, 0.5): 
\begin{verbatim}
> gamma.fun = ellipse(a = 0.35, b = 0.25)
> bin.obs = par2obs(m = 100, pi.in = 0.6, pi.out = 0.4, design = 'J',
+ gamma.fun, center = c(0.5, 0.5))
\end{verbatim}
Similarly, the following code generates a $100 \times 100$ Gaussian-noised image with a triangle boundary using a random uniform design with a reference point of (0.5, 0.5):
\begin{verbatim}
> gamma.fun = triangle2(0.5)
> norm.obs = parnormobs(m = 100, mu.in = 1, mu.out = -1, sd.in = 1, 
+ sd.out = 1, design = 'U', gamma.fun, center = c(0.5, 0.5))
\end{verbatim}           
The output of either \code{par2obs} or \code{parnormobs} is a list containing the intensities $Y$ in a vector named \code{intensity}, the vectors \code{theta.obs} and \code{r.obs} containing the polar coordinates of the pixel locations $X$, and the reference point contained in a vector named \code{center}.  Image data to be analyzed using \pkg{BayesBD} can be in a list of this form, or may be a \code{.png} or \code{.jpg} image file.  

\begin{table}[!ht]
	\centering
	\begin{tabular}{ll}
		\toprule
		& \code{par2obs} arguments                                                                                                                                                                                                               \\ \midrule
		\code{m}                       & Generate $m\times m$ observations over the unit square.                                                                                                                                                                                           \\
		\code{pi.in}                   & The success probability, $P(Y_i = 1)$, if $X_i\in I_1$.                                                                                                                                                                                     \\
		\code{pi.out}                  & The success probability, $P(Y_i = 1)$, if $X_i\in I_2$.                                                                                                                                                                                     \\
		\multirow{3}{*}{\code{design}} & \multirow{3}{*}{\begin{tabular}[c]{@{}l@{}l@{}}Determines how locations $X_i$ are determined: 'D' for deterministic\\  (equally-spaced grid) design, 'U' for completely uniformly random, \\ or 'J' for jitteredly random design.\end{tabular}} \\
		&                                     \\                                                                                                                                                                                                        \\
		\multirow{2}{*}{\code{center}} & \multirow{2}{*}{\begin{tabular}[c]{@{}l@{}}Two-dimensional vector of Euclidean coordinates (x,y) of the reference\\  point in $I_1$.\end{tabular}}                                                                                          \\
		&                                                                                                                                                                                                                                             \\
		\code{gamma.fun}              & A function to generate boundaries, e.g. ellipse or triangle2.                                                                                                                                                                               \\ \bottomrule
	\end{tabular}
	\caption{Arguments of the \code{par2obs} function. \label{tbl:par2obs}}
\end{table}

\subsection{Analysis and visualization} 
There are two functions to draw posterior samples following Algorithm 1 and 2 based on images either simulated or provided by users: \code{fitBinImage} for binary images, and \code{fitContImage} for Gaussian-noised images. These sampling functions take the same arguments, with the exception of the \code{ordering} input which is duplicated in \code{fitContImage} to allow ordering of the two parameters, i.e., the mean and standard deviation. The inputs for \code{fitBinImage} are summarized in Table~\ref{tbl:bayesbd}.  We have included a function \code{rectToPolar} to facilitate formatting the image data for \code{fitBinImage} and \code{fitContImage} by converting the rectangular coordinates of the pixels to polar coordinates.  The initial boundary is a circle with radius \code{inimean} and center \code{center}.  The radius \code{inimean} may be specified by the user or left blank, in which case it will be estimated using maximum likelihood. 

\begin{table}[!ht]
	\centering
	\begin{tabular}{ll}
		\toprule
		& \code{fitBinImage} arguments                                                                                                                                                                                                 \\ \midrule
\code{image} & \begin{tabular}[c]{@{}l@{}}
The noisy observation, either a list with elements: 
\\ \quad  \texttt{intensity}, a vector of intensities; 
\\  \quad \code{theta.obs} a vector of pixel radian measure from \code{center};
\\ \quad \code{r.obs} a vector of pixel radius measure from \code{center}; \\
or a string giving the path to a \code{.png} or \code{.jpg} file. \end{tabular}\\
\code{gamma.fun} & the true boundary, if known, used for plotting. \\
\code{center} & the reference point in Euclidean coordinates. \\
\code{inimean} & A constant specifying the initial boundary $\mu$. Defaults to NULL, \\
&in which case $\mu$ is estimated automatically using maximum likelihood.\\
\code{nrun} & The number of MCMC runs to keep for estimation. \\
\code{nburn} & The number of initial MCMC runs to discard. 	\\
\code{J}    & The number of eigenfunctions to use in estimation is $2J+1$. \\
\code{ordering}    &\begin{tabular}[c]{@{}l@{}} Indicates which Bernoulli distribution has higher success probability: 
			\\ \quad \texttt{"I"}, the Bernoulli distribution inside the boundary; \\
				\quad \texttt{"O"}, ther Bernoulli distribution outside the boundary; \\
				\quad \texttt{"N"},  no ordering information is available.   \end{tabular}\\
\code{mask} & \begin{tabular}[c]{@{}l@{}}
Logical vector (same length as \texttt{obs}\$\texttt{intensity}) to indicate region of interest.\\
Should this data point be included in the analysis?  Defaults to NULL\\
and uses all data points.  \end{tabular}\\
\code{slice} & \begin{tabular}[c]{@{}l@{}}
Should slice sampling be used to sample Fourier basis function coefficients?\end{tabular}\\
\code{outputAll} & Should all posterior samples be returned?  \\ \bottomrule	\end{tabular}
	\caption{Arguments of \code{fitBinImage}.}
	\label{tbl:bayesbd}
\end{table}

If argument \code{outputAll} is \code{FALSE}, the functions \code{fitBinImage} and \code{fitContImage} produce output vectors \code{theta}, \code{estimate}, \code{upper}, and \code{lower} giving a grid of 200 values on $[0,2\pi]$ on which the boundary $\gamma$ is estimated, along with $95\%$ uniform credible bands.  If argument \code{outputAll} is \code{TRUE}, the functions also return posterior samples of $(\xi, \rho)$ and Fourier basis function coefficients $z$.

Following the examples of data simulation for binary and Gaussian-noised images in the previous section, we can obtain posterior samples via\\

\begin{verbatim}> bin.samples= fitBinImage(bin.obs, c(.5,.5), NULL, 4000, 1000, 10, "I", NULL, TRUE,                    
+ FALSE)
\end{verbatim} 
for a binary image and
\begin{verbatim}
> norm.samples = fitContImage(norm.obs, c(.5,.5), NULL, 4000, 1000, 10, "I", "N", NULL,
TRUE, FALSE)
\end{verbatim}
for a Gaussian-noised image.  For each sampling function, we have set the center of the image as $(0.5, 0.5)$, instructed the sampler to use a mean function of $\mu(\cdot) = 0.4$, keep $4000$ samples, burn $1000$ samples, use $L = 2\times 10+1 = 21$ basis functions to model $\gamma$, use slice sampling for the basis function coefficients, and return only the plotting results.

\begin{figure}[!ht]
	\centering
	\includegraphics[width=0.7\textwidth, trim={2cm 3cm 2cm 3cm},clip]{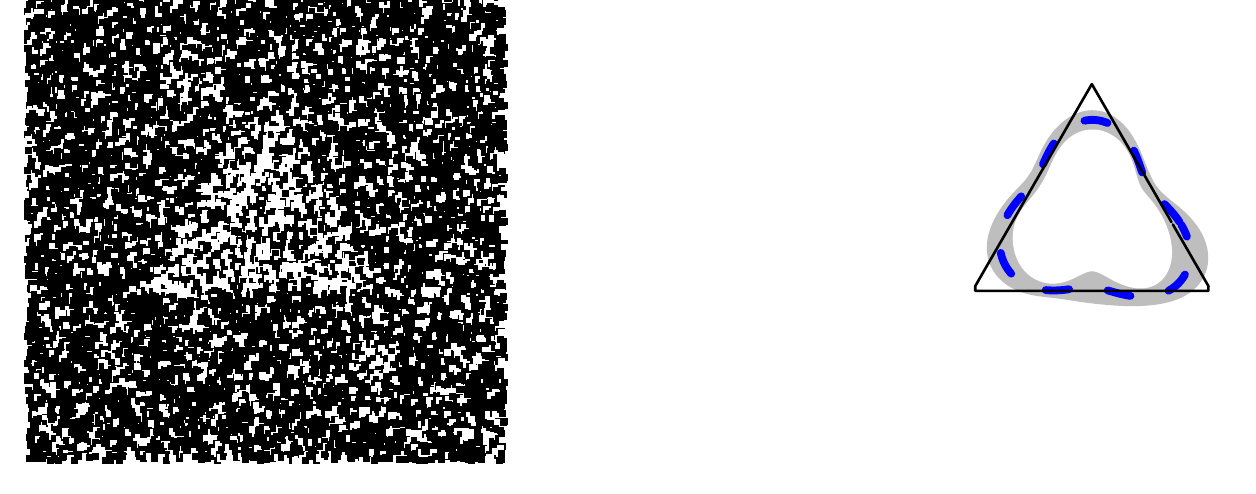}
	\caption{Output from \code{fitBinImage} as plotted using \code{plotBD}. The plot on the left is the simulated binary observation, and the plot on the right is the estimated boundary and 95\% uniform credible bands.}
	\label{plot:plotBD}
\end{figure}

Using the function \code{plotBD} we can easily construct plots of our model results.  There are two arguments to \code{plotBD}: \code{fitted.image} is a list of results from \code{fitBinImage} or \code{fitContImage}, and \code{plot.type} which indicates whether to plot the image data only, the posterior mean and credible bands, or the posterior mean overlaid on the image data.  Using the posterior samples we obtained from the Gaussian-noised image above, we can plot our results with the following code
\begin{verbatim}
> par(mfrow = c(1,2))
> plotBD(norm.samples, 1)
> plotBD(norm.samples, 2)
\end{verbatim}     
to produce Figure~\ref{plot:plotBD}. 

\section{BayesBD \pkg{shiny} app}
In order to reach a broad audience of users, including those who may not be familiar with \code{R}, we have created a \pkg{shiny} app version of \pkg{BayesBD} to implement the boundary detection method. The app has the full functionality to reproduce the simulations in Section~\ref{S:4} and conduct real data applications using images uploaded by users. 
The app is accessible both from the package by running the code \code{BayesBDshiny()} in an \code{R} session or externally by visiting \url{https://syring.shinyapps.io/shiny/}.  With the app users can analyze real data or produce a variety of simulations.

\begin{figure}[!ht]
	\centering
	\includegraphics[width=.90\textwidth]{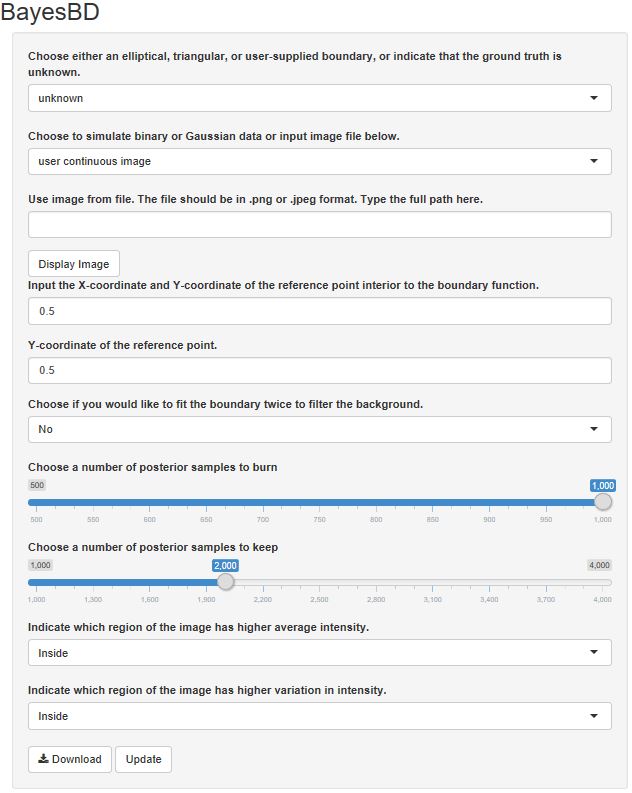}
	\caption{Screenshot of shiny app implementing \pkg{BayesBD}.}
	\label{fig:shiny_disp}
\end{figure}

Once the app is open, users are presented with an array of inputs to set as illustrated in Figure~\ref{fig:shiny_disp}.  In order to analyze real data, the user should select "user continuous image" or "user binary image" from the second drop-down menu.  Next, the user inputs the system path to the \code{.png} or \code{.jpg} file.  A plot of the image will appear and the user will be prompted to identify the image center with a mouse click.  The user has some control over the posterior sample size, but we recommend to first limit the number of available samples in order to display results quickly. Finally, the user may enter ordering information for the mean and variance of pixel intensities at the bottom of the display.

For simulations, we first select either an elliptical or triangular boundary, or upload an Rscript with a custom boundary function.  Next, the user instructs the app to either simulate a binary or Gaussian-noised image, or to use binary or Gaussian data the user has uploaded.  In addition to the boundary function, the user specifies the reference point.  Sample sizes for simulations are kept at $100 \times 100$ pixels.  Finally, the last several inputs allow the user to customize the intensity parameters $(\xi, \rho)$ for binary and Gaussian simulations.  Once the user has selected all settings, clicking the "Update" button at the bottom of the window will run the posterior sampling algorithm, which should take less than a minute on a typical computer for image data of $100 \times 100$ pixels.  The "Download" button provides the user with a file indicating which pixels were contained inside the estimated boundary as determined by the outer edge of the $95\%$ uniform credible bands.      

\section{Performance tests}
\label{S:4}
\subsection{Comparison of sampling methods}

The main aim of this section is to highlight the speed improvements we have made in the latest version of \pkg{BayesBD}. Our flexibility in choosing between slice and Metropolis-Hastings (MH) sampling algorithms gives users the potential to unlock efficiency gains. We highlight these gains below for both binary and Gaussian-noised simulations, and note that the faster MH method suffers little in accuracy.  

In our performance tests, we consider the following examples, which correspond to examples B2, B3, and G1 from \citet{Li.Ghosal.2015}.  
\begin{enumerate}
	\item[S1. ]  Image is an ellipse centered at $(0.1,0.1)$ and rotated $60\degree$ counterclockwise.  Intensities are binary with $\pi_1 = 0.5$ and $\pi_2 = 0.2$.  
	\item[S2. ]  Image is an equilateral triangle centered at $(0,0)$ with height $0.5$.  Intensities are binary with $\pi_1 = 0.5$ and $\pi_2 = 0.2$.  
	\item[S3. ]  Image is an ellipse centered at $(0.1,0.1)$ and rotated $60\degree$ counterclockwise.  Intensities are Gaussian with $\mu_1 = 4$, $\sigma_1 = 1.5$, $\mu_2 = 1$, and $\sigma_2 = 1$. 
\end{enumerate}

We simulated each case $100$ times using $n = 100 \times 100$ observations per simulation.  In each run, we sampled 4000 times from the posterior after a 1000 sample burn in.  The results of our performance tests comparing slice with MH sampling are summarized in Table~\ref{tbl:sample}.  

If Metropolis-Hastings sampling is used instead of slice sampling, we observe a speed up by about a factor of two for binary images, seven for the Gaussian-noised image.  Slice sampling is guaranteed to produce unique posterior samples, and may give better results than Metropolis-Hastings samplers, especially when Metropolis-Hastings mixes poorly producing many repeated samples.  However, slice sampling may involve a very large number of proposed samples for each accepted sample, requiring many likelihood evaluations.  On the other hand, each Metropolis Hastings sample requires only two evaluations of the likelihood.  

To measure the accuracy of \pkg{BayesBD} we use three metrics: the Lebesgue error, which is simply the area of the symmetric difference between the posterior mean boundary and the true boundary; the Dice Similarity Coefficient(DSC), see \citet{mritc}; and Hausdorff distance, see \citet{RadOnc} and \citet{RadOncR}.  We have included the utility functions \code{lebesgueError}, \code{dsmError}, and \code{hausdorffError}, which take as input the output of either \code{fitBinImage} or \code{fitContImage} and output the corresponding error.  In the binary image examples considered, the different sampling algorithms did not affect the accuracy of the posterior mean boundary estimates when measured by Lebesgue error, i.e., the area of the symmetric difference between the estimated boundary and the true boundary.  For the Gaussian-noised image, the slice sampling method produced Lebesgue errors approximately an order of magnitude smaller than when using Metropolis-Hastings sampling, but the overall size of the errors was still small in both cases and practically indistinguishable when plotting results.  With our built-in functions, it is easy to reproduce Example S2 in Table~\ref{tbl:sample} with the following code:
\begin{verbatim}
> gamma.fun = triangle2(0.5)
> for(i in 1:100){
+ 	norm.obs = par2obs(m = 100, pi.in = 0.5, pi.out = 0.2, design = 'J',
+		 			   center = c(0.5,0.5), gamma.fun)
+ 	norm.samp.MH = fitBinImage(norm.obs, gamma.fun,NULL,NULL, 4000, 1000,
+		 						   10,"I",rep(1,10000), FALSE, FALSE)
+ 	norm.samp.slice = fitBinImage(norm.obs, gamma.fun,NULL,NULL, 4000, 
+		 			  						1000, 10,"I",rep(1,10000), TRUE, FALSE)
+ 	print(c(dsmError(norm.samp.MH), hausdorffError(norm.samp.MH), 	
+ 			  	lebesgueError(norm.samp.MH), dsmError(norm.samp.slice), 
+ 			  	hausdorffError(norm.samp.slice), lebesgueError(norm.samp.slice)))
+ }
\end{verbatim}

\begin{table}[!ht]
	\centering
	\begin{tabular}{cccccc}\toprule
		\multicolumn{1}{c}{Example} & \multicolumn{1}{c}{Sampling Method} & \multicolumn{1}{c}{Runtime (s)} & \multicolumn{1}{c}{Lebesgue Error}& \multicolumn{1}{c}{DSC}& \multicolumn{1}{c}{Hausdorff} \\ \midrule
		\multirow{2}{*}{S1}         & MH                                  & 58                              & 0.01(0.01)   & 0.02(0.00) & 0.02(0.00)                     \\
		& Slice                               & 100                             & 0.01(0.00)      & 0.02(0.00) & 0.02(0.00)                    \\
		\multirow{2}{*}{S2}         & MH                                  & 45                              & 0.02(0.00)  & 0.09(0.02) & 0.09(0.01)                       \\
		& Slice                               & 82                              & 0.02(0.00)             & 0.09(0.01) & 0.09(0.01)              \\
		\multirow{2}{*}{S3}         & MH                                  & 66                              & 0.01(0.01)     & 0.01(0.01) & 0.01(0.01)                       \\
		& Slice                               & 488                             & 0.00(0.00)           & 0.01(0.01) & 0.01(0.01)              \\ \bottomrule
	\end{tabular}
	\caption{Average Runtimes (in seconds) and Lebesgue errors (with standard deviations) of posterior mean boundary estimates using \code{c++}.}
	\label{tbl:sample}
\end{table}

\subsection{Comparison of coding platforms}

Our use of \code{c++} for posterior sampling has led to very significant efficiency gains over using R alone.  Implementation of \code{c++} with R is streamlined using \pkg{Rcpp} and \pkg{RcppArmadillo}.

The first implementation of \pkg{BayesBD} (version 0.1 in~\cite{Li.BD.2015}) was entirely written in \code{R}.  The results in Table~\ref{tbl:code} labeled \code{R} Code reflect this first version of the package, while those labeled \code{c++} Code are from the new version.  The main takeaway is that the new code developed using \code{c++} is at least three times faster in the binary image examples and over six times faster for the Gaussian-noised image example, both while using slice sampling. 

\begin{table}[!ht] 
	\centering
	\begin{tabular}{ccc}\toprule
		\multicolumn{1}{l}{Example} & \multicolumn{1}{l}{Coding Method} & \multicolumn{1}{l}{Runtime (s)} \\ \midrule
		\multirow{2}{*}{S1}         & \code{c++}                    & 100                             \\
		& \code{R}                                 & 375                             \\
		\multirow{2}{*}{S2}         & \code{c++}                   & 82                              \\
		& \code{R}                            & 246                             \\
		\multirow{2}{*}{S3}         & \code{c++}                    & 488                             \\
		& \code{R}                               & 3327                            \\ \bottomrule
	\end{tabular}
	\caption{Average Runtimes (in seconds) using slice sampling.}
	\label{tbl:code}
\end{table} 

\subsection{Scalability of \pkg{BayesBD}}
\begin{figure}
	\centering
		\includegraphics[width=0.4\textwidth]{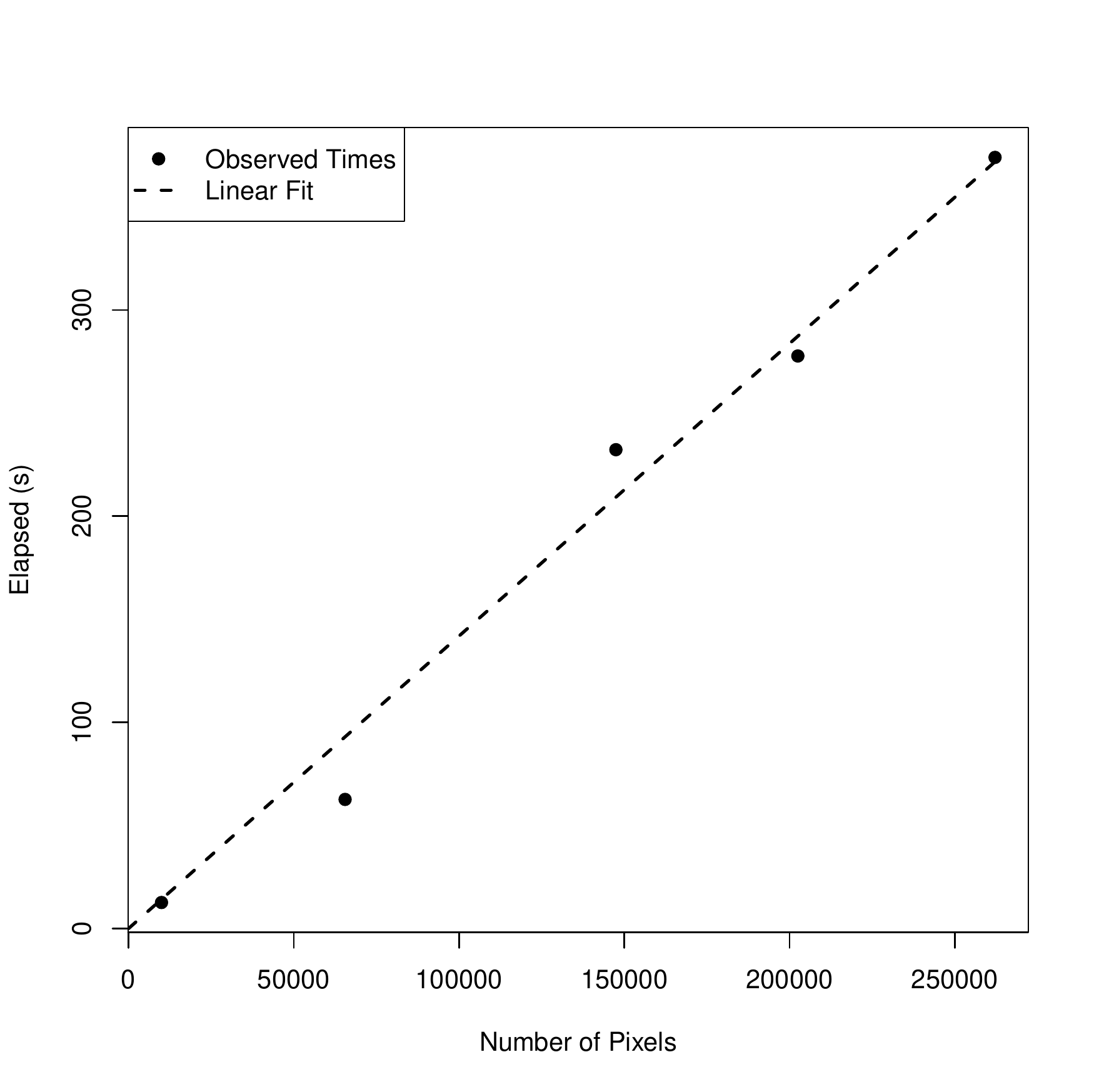}
		\includegraphics[width=0.4\textwidth]{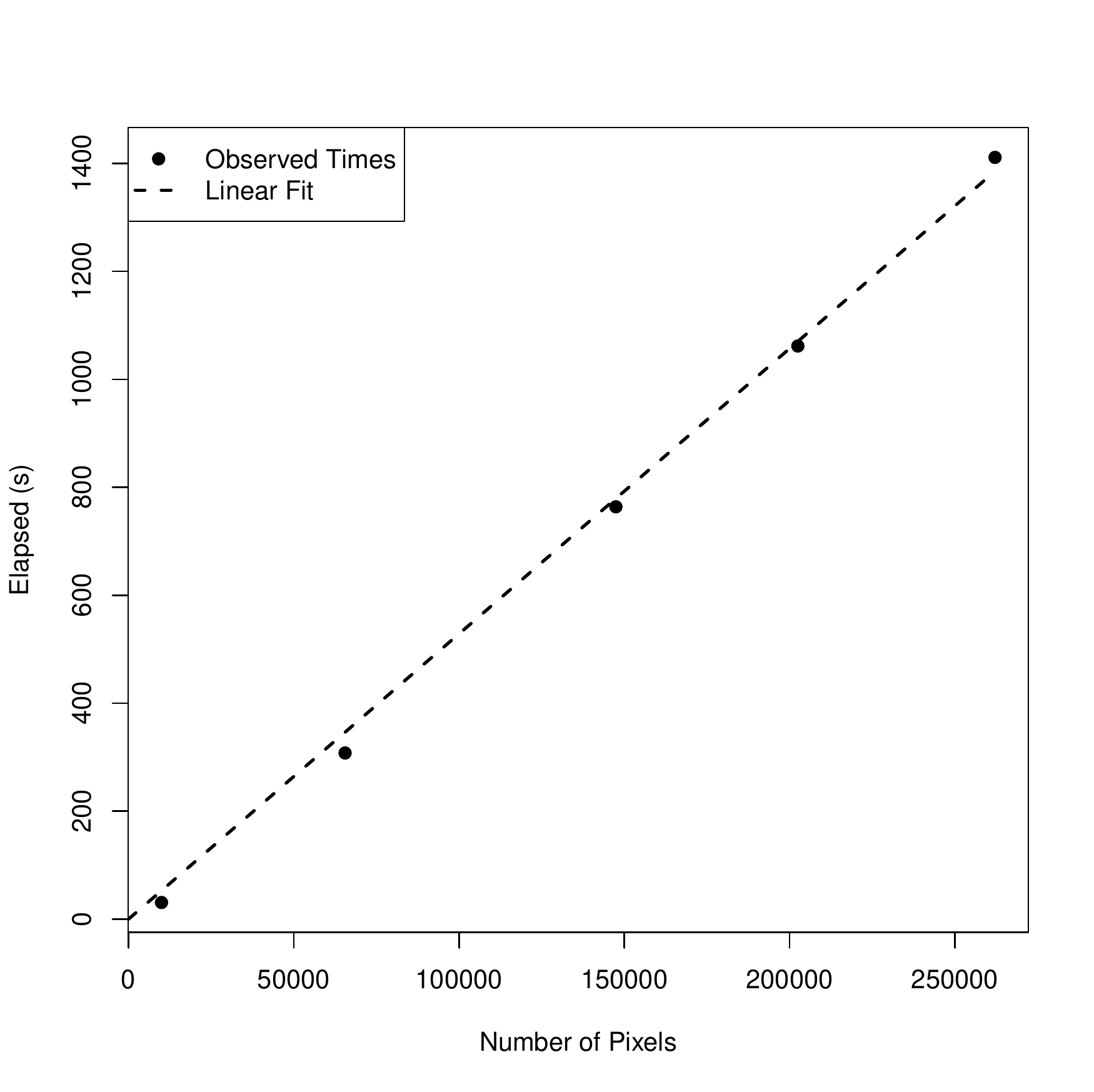}
	\caption{Left: Runtimes of \code{fitBinImage}. Right: Runtimes of \code{fitContImage} }
	\label{fig:complexity}
\end{figure}

		The Gaussian process is notorious for scaling poorly as $n$ increases because it is usually necessary to invert of a large covariance matrix. By utilizing the analytical decomposition of a GP kernel in~\eqref{eq:decomposition}, we eliminates the step of inverting a $n$ by $n$ covariance matrix and the \pkg{BayesBD} package appears to achieve a linear complexity. 
		We investigate the scalability of \pkg{BayesBD} by plotting the system time against sample size for \code{fitBinImage} and \code{fitContImage} in Figure~\ref{fig:complexity} using a triangular boundary curve and $5000$ MCMC iterations. Both algorithms appear to scale approximately linearly in number of pixels, which makes sense as the costliest computations in Steps 1 and 3 in Algorithms 1 and 2 only involve sums over the $n$ pixels. 

\subsection{Comparison with existing packages}

\begin{figure}[htbp]
	\centering
		\includegraphics[width=1.00\textwidth]{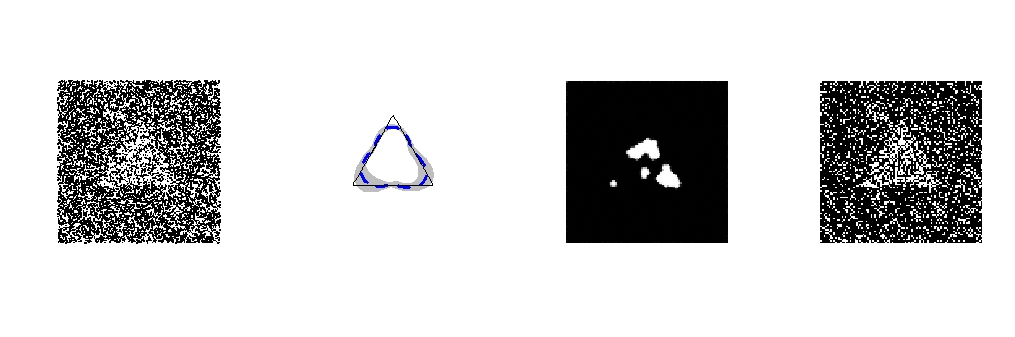}
				\includegraphics[width=1.00\textwidth]{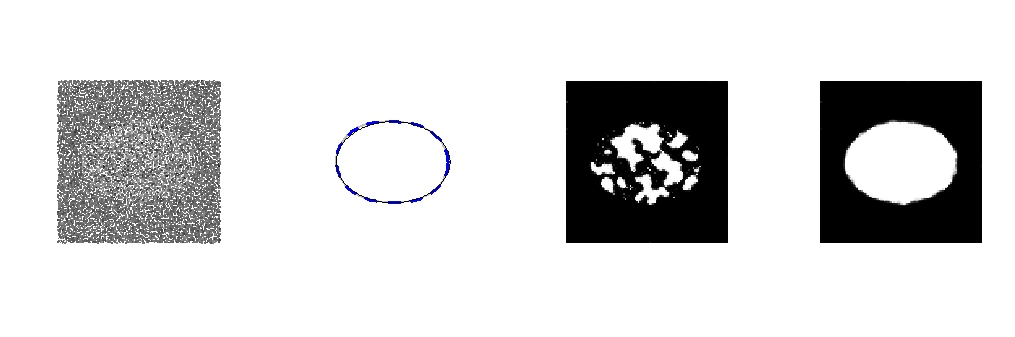}
	\caption{Left to right: image, \pkg{BayesBD} boundary estimate, \pkg{mritc} filtered image, and \pkg{bayesImageS} filtered image.}
	\label{fig:compare}
\end{figure}

Although no packages besides \pkg{BayesBD} provide boundary estimation, there are several existing packages that can provide image segmentation or filtering.  Below we make some qualitative comparisons between \pkg{BayesBD} and \pkg{mritc}, \pkg{bayesImageS}, and \pkg{bayess}; see \citet{mritc}, \citet{bayesImageS}, and \citet{bayess} in a later section.  Figure~\ref{fig:compare} compares these packages using two simulated images with Bernoulli and Gaussian noise, respectively.  \pkg{BayesBD} gives very reasonable estimates for the true boundaries; \pkg{mritc} package fails to deliver a recognizable smoothed image in either example; and \pkg{bayesImageS} was able to produce a very clear segmentation for the Gaussian-noised ellipse example, but not for the triangle image with Bernoulli noise.     

\section{Real data application}

\subsection{Medical imaging}

\begin{figure}[!ht]
	\centering
	\includegraphics[width=0.60\textwidth]{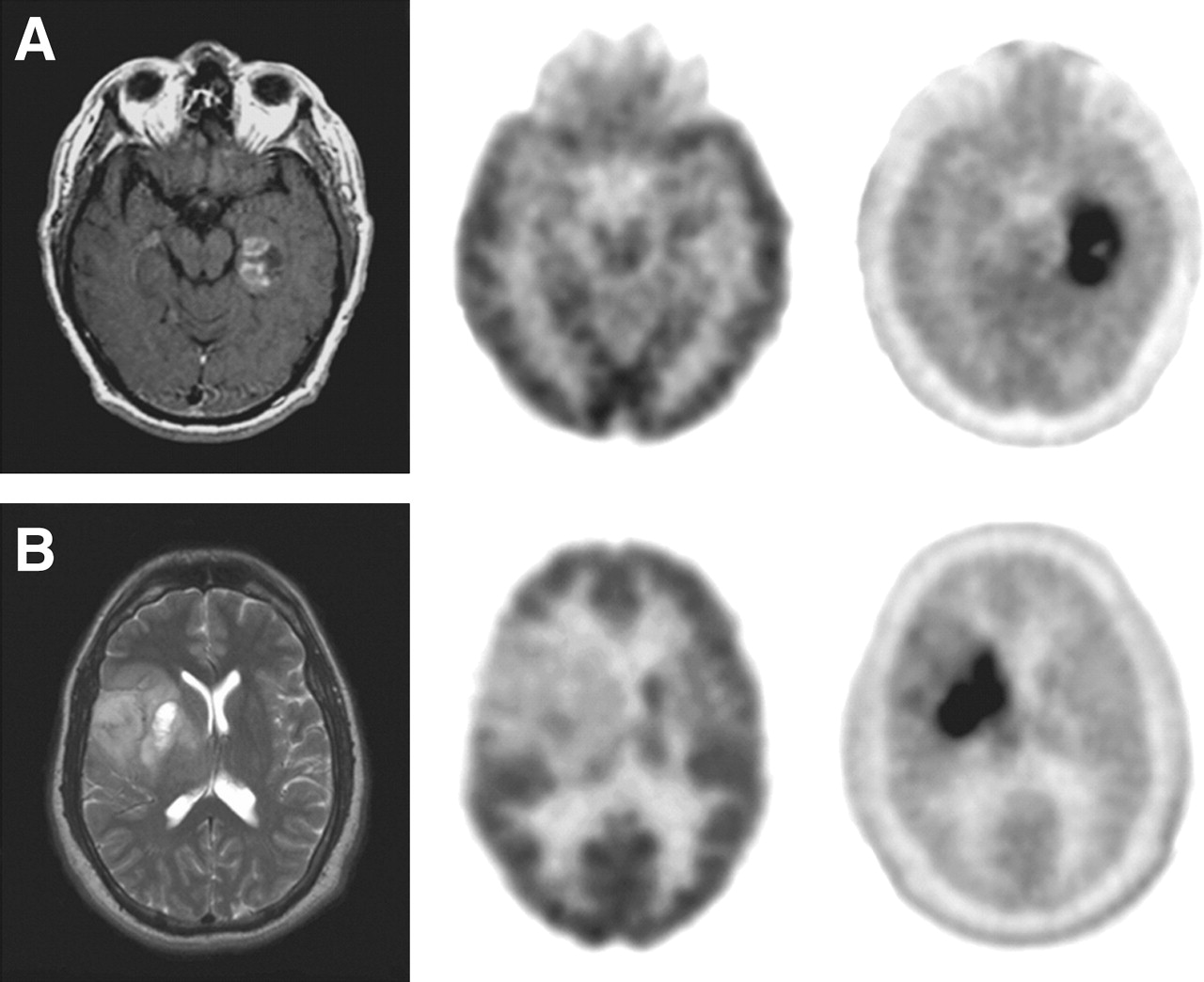}
	\caption{MRI (left), F-FDG PET (middle), and F-FDOPA PET (right) of glioblastoma (A) and grade II oligodendroglioma (B).  Image taken from \citet{PET}.}
	\label{fig:pet}
\end{figure}

\citet{PET} studied the performance of two different tracers used in conjunction with positron emission tomography (PET) scans in identifying brain tumors. Figure~\ref{fig:pet}, reproduced from~\citep{PET}, gives an example of the image data used in diagnosing tumors, and demonstrates their conclusion that the F-FDOPA tracer provides a more informative PET scan than the F-FDA tracer. We use the \pkg{BayesBD} package to analyze the F-FDOPA PET scan images in Figure~\ref{fig:pet}.  The tumor imaging data along with sample code for reproducing the following analysis can be found in the documentation to the \pkg{BayesBD} package. 

We convert the two F-FDOPA PET images in Figure~\ref{fig:pet} into $111\times 111$-pixel images and normalize the intensities to the interval [0,10].  The pixel coordinates are a grid on $[0,1]\times[0,1]$ and we choose reference points $(0.7,0.5)$ and $(0.4,0.55)$ for each image, roughly corresponding to the center of the darkest part of each image. 
We use the default mean function, choose $J=10$ for $21$ basis functions, and sample $4000$ times after a $1000$ burn-in using MH sampling.

\begin{figure}[!ht]
	\begin{tabular}{cc}
		\centering
		\includegraphics[width=0.48\textwidth, trim = {0cm 0cm 0cm 0cm},clip]{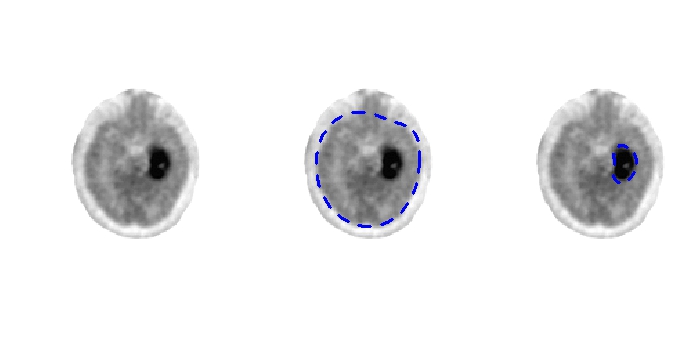} & 
		\includegraphics[width=0.48\textwidth, trim = {0cm 0cm 0cm 0cm},clip]{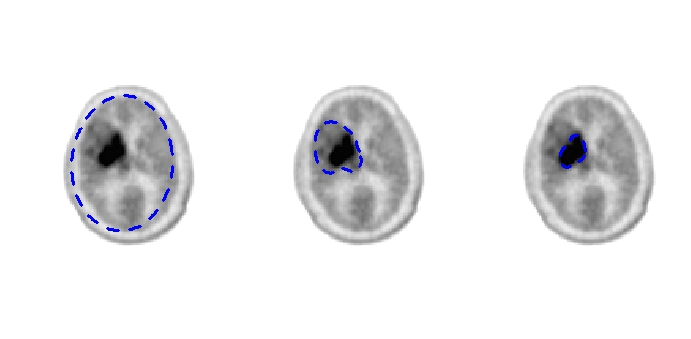} \\
		A. Glioblastoma & B. Grade II oligodendroglioma 
	\end{tabular} 
	\caption{F-FDOPA PET images from \citet{PET} (left) fit twice, and (right) fit three times to filter the background and find features at increasing granularity.}
	\label{fig:pet_bd}
\end{figure}
Figure~\ref{fig:pet_bd} displays posterior mean boundary estimates for the F-FDOPA images in Figure~\ref{fig:pet}.  From the analysis on glioblastoma (A) in the first two plots, it seems that 
that we accurately capture the regions of interest in the F-FDOPA PET images. Furthermore, it is expected that the Gaussian assumption on the real data may fail, and this shows that the method implemented in \pkg{BayesBD} is robust to model misspecifications, thus practically useful.

Tumor heterogeneity, which is not unusual in many applications, may make the boundary detection problem more challenging~\citep{Heppner:84, ananda2012automatic}. The \pkg{BayesBD} package allows us to address tumor heterogeneity by a repeated implementation. We first apply \code{fitContImage} to the entire image, which includes a white background not of interest, and produce the estimated boundary.  This step succeeds in separating the brain scan from the white background.  A second run is performed on the subset of the image inside the outer 95\% uniform credible band, producing a nested boundary. In general, this technique can be used in a multiple region setting where the data displays more heterogeneity than the simple "image and background" setup in \eqref{eq:setup}.     

\subsection{Satellite imaging}

\begin{figure}[htbp]
	\centering
						\includegraphics[width=1.00\textwidth]{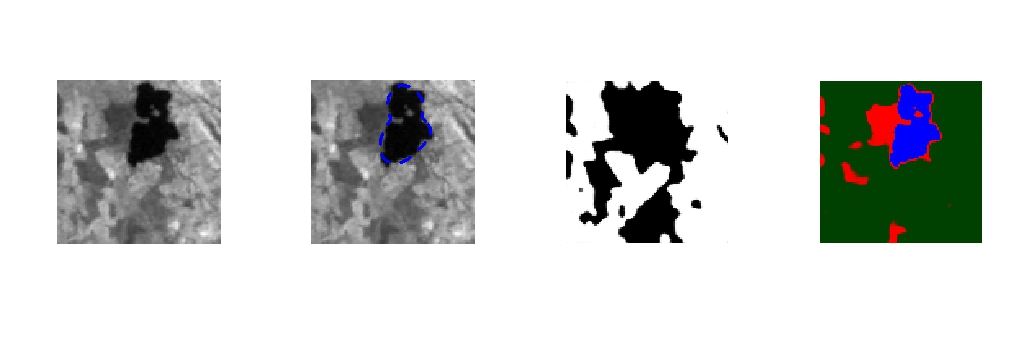}
	\caption{Left to right: image, \pkg{BayesBD} boundary estimate, \pkg{mritc} filtered image, and \pkg{bayess} segmented image.}
	\label{fig:compare_lake}
\end{figure}

We compared the performance of \pkg{BayesBD} with the \code{R} packages \pkg{mritc} and \pkg{bayess} using an image of Lake Menteith available in the \pkg{bayess} package.  \pkg{BayesBD} gives a very reasonable estimate for the boundary of the lake even though it is not smooth.  The \pkg{mritc} package again does not provide useful output in this example, but \pkg{bayess} produces a nicely-segmented image; see Figure~\ref{fig:compare_lake}.

\section{Summary}
\label{S:discuss}

\pkg{BayesBD} is a new computational platform for image boundary analysis with many advantages over existing software.  The underlying methods in functions \code{fitBinImage} and \code{fitContImage} are based on theoretical results ensuring their dependability over a range of problems.  Our use of \pkg{Rcpp} and \pkg{RcppArmadillo} help make \pkg{BayesBD} much faster than base \code{R} code and further speed can be gained by our flexible sampling algorithms.  Finally, our integration with \pkg{shiny} provides users with an easy way to utilize our package without having to code. 

For the latest updates to \pkg{BayesBD} and requests, readers are recommended to check out the package page at CRAN or refer to the Github page at \url{https://github.com/nasyring/GSOC-BayesBD}.          

\section{Acknowledgments} 
This work was partially supported by the Google Summer of Code program. We thank the Associate Editor and one anonymous referee for comprehensive and constructive comments that helped to improve the paper. 

\bibliography{syring-li}

\address{Nicholas Syring\\
	Department of Statistics\\
	North Carolina State University, 5109 SAS Hall\\
	2311 Stinson Dr.\\ 
	Raleigh, NC 27695 USA\\}
\email{nasyring@ncsu.edu}

\address{Meng Li\\
	Rice University\\
	Department of Statistics\\
	6100 Main St\\
	Houston, TX 77251-1892 USA\\}
\email{meng@rice.edu}

\end{article}

\end{document}